\documentclass{emulateapj}
\usepackage{natbib}

\newcommand{\msun}{\ensuremath{{\rm M}_\odot}}
\newcommand{\psr}{PSR\,J0751+1807}

\begin{document}

\journalinfo{Submitted to the Astrophysical Journal}

\title{A 2.1 Solar Mass Pulsar Measured by Relativistic Orbital Decay}

\author{David J. Nice and Eric M. Splaver}
\affil{Physics Department, Princeton University \\ 
Princeton, NJ  08544}

\author{Ingrid H. Stairs}
\affil{Department of Physics and Astronomy, University of British
Columbia\\
6224 Agricultural Road, Vancouver, BC V6T 1Z1, Canada}

\author{Oliver L\"ohmer and Axel Jessner}
\affil{Max-Planck-Institut f\"ur Radioastronomie \\
Auf dem H\"ugel 69, D-53121 Bonn, Germany }

\author{Michael Kramer}
\affil{University of Manchester, Jodrell Bank Observatory \\
Macclesfield, Cheshire, SK11 9DL, UK}

\and

\author{James M. Cordes}
\affil{Astronomy Department and NAIC, Cornell University \\
Ithaca, NY 14853}

\medskip

\submitted{Submitted to the Astrophysical Journal, 15 June 2005;  Revised, 31 July 2005}

\begin{abstract}

\psr\ is a millisecond pulsar in a circular 6\,hr binary system with a
helium white dwarf secondary.  Through high precision pulse timing
measurements with the Arecibo and Effelsberg radio telescopes, 
we have detected the decay of its orbit due to emission
of gravitational radiation.  This is the first detection of the
relativistic orbital decay of a low-mass, circular binary pulsar
system.  The measured rate of change in orbital period, corrected
for acceleration biases, is $\dot{P_b}=(-6.4\pm 0.9)\times 10^{-14}$.  
Interpreted in the context of general
relativity, and combined with measurement of Shapiro delay, it implies
a pulsar mass of $2.1\pm 0.2$\,\msun, the most massive pulsar measured.
This adds to the emerging trend toward relatively high neutron star
masses in neutron star--white dwarf binaries.  Additionally, there
is some evidence
for an inverse correlation between pulsar mass and orbital period in
these systems.  We consider alternatives to the general relativistic
analysis of the data, and we use the pulsar timing data to place
limits on violations of the strong equivalence principle.

\end{abstract}

\keywords{gravitation---binaries: general---pulsars: individual
(PSR~0751+1807)}

\section{Introduction}\label{sec:intro}

A fundamental prediction of general relativity is the emission of
gravitational radiation by binary systems.  The consequent loss of
energy and angular momentum from these systems causes their orbital
periods to decrease \citep{pet64}.  Previous to the present work,
relativistic orbit decay had been detected in five binary pulsar
systems.  Four of these systems consist of a mildly recycled pulsar
bound to a second neutron star in an eccentric orbit
\citep{tw89,sttw02,dk96,kll+05}.  The fifth system, that of PSR J1141$-$6545,
contains a young
pulsar and a $\sim$1\,\msun\ white dwarf in an eccentric orbit
\citep{bokh03}.

In this paper, we report the measurement of the relativistic decay of
\psr, a millisecond pulsar in a 6\,hr orbit \citep{lzc95}.  This
system differs in several ways from the other binaries in which relativistic
decay has been detected.  First, its rotation period of  3.4\,ms
is an order of magnitude smaller than the rotation periods of the
pulsars in the other systems, implying greater accumulation of 
angular momentum during its binary accretion phase.  Second, its magnetic field
is substantially lower than those of the other systems, presumably
because of differences in orbital evolution and accretion \citep{bha02b}.
Third, its orbit is extremely circular, with eccentricity under $2\times
10^{-6}$, a consequence of tidal circularization during the late
stages of the secondary \citep{phi92b}.  Fourth, the secondary to \psr\ is a 
low mass helium white dwarf, several times lighter than the secondary stars in
the other systems.

\begin{deluxetable*}{llcccccc}[t]
\tablewidth{0pt}
\tablecaption{Summary of Observations\label{tab:obs}}
\tablehead{
\multicolumn{1}{c}{Observatory}   
              & \multicolumn{1}{c}{System}   
                         & Dates         & Frequency & Bandwidth  & Number & Typical        & RMS         \\
              &          &                &   (MHz)  &  (MHz)     & of  & Integration    & Residual    \\
              &          &                &          &            & TOAs & (min)          & ($\mu$s)    \\
} 
\startdata
Arecibo       & Mark III & 1993.8--1994.4 & \phn 430 &      \phn 8&    1190 &   0.5        &      25     \\
              & Mark IV  & 1998.9--2004.1 & \phn 430 &       10   &    1007 &   3.2        &  \phn 5     \\
              & Mark IV  & 1997.9--2001.5 & 1410     &       10   & \phn 109 &   3.2        &  \phn 7     \\
              & WAPP     & 2004.1--2004.1 & 1145     &       50   & \phn 562 &   0.5        &      10     \\
              & WAPP     & 2004.1--2004.1 & 1195     &       50   & \phn 559 &   0.5        &  \phn 8     \\
              & WAPP     & 2004.1--2004.1 & 1395     &       50   & \phn 559 &   0.5        &  \phn 8     \\
              & WAPP     & 2004.1--2004.1 & 1445     &       50   & \phn 559 &   0.5        &  \phn 8     \\
Effelsberg    & EBPP     & 1997.0--2004.6 & 1409     &       56   & \phn 490 &   7.0        &  \phn 7     \\
\enddata
\end{deluxetable*}

Measurements of relativistic orbital phenomena in binary systems yield
constraints on the masses of the stellar components of the systems.
Neutron star mass measurements, in turn, serve as probes of the properties of
nuclear matter at high densities \citep[e.g.,][]{lp01}.  
The maximum achievable neutron
star mass depends on the equation of state of nuclear matter.  Soft
equations of state, expected if the core of the neutron star is
composed of non-nucleonic matter, predict the maximum neutron star
mass $\lesssim$2\,\msun, while stiffer equations of state allow higher
values.

Reviewing the field several years ago, \cite{tc99} found all radio pulsar mass
measurements to be consistent with a narrow underlying Gaussian mass distribution,
$1.35\pm 0.04$\,\msun.  More recent observations 
have found (1) neutron star masses in neutron star--neutron star binary 
systems lie in the range 1.18 to 1.44\,\msun;  (2) PSR~J1141$-$6454,
a young pulsar in an eccentric orbit with a white dwarf secondary, 
has mass 1.30\,\msun; and (3) neutron stars in circular neutron star--wide 
dwarf systems may have somewhat larger masses, but the uncertainties are large
\citep[e.g.,][]{nss05}.

Neutron star masses can also be measured in X-ray binary systems.  
Spectroscopic and eclipse observations typically 
yield neutron star mass measurements  in the range 1.1 to 1.5\,\msun,
but there is evidence for a few more massive neutron stars:
$1.78\pm 0.2$\,\msun\ for the neutron star in Cygnus X-2 \citep{ok99},
$2.44\pm 0.27$\,\msun\ for the neutron star in 4U1700$-$37 \citep{cgc+02}, and
$2.27\pm 0.17$\,\msun\ for the neutron star in Vela X-1 \citep{qna+03}.
See the review papers by \cite{lp04} and
\cite{cc05} for further references.

We undertook high precision pulse timing observations of \psr\ 
with the goal of measuring its orbital decay, both to test the expected 
relativistic
behavior of its binary system and to deduce the pulsar and secondary
star masses.  In \S\ref{sec:obs} we describe the observations in
detail.  In \S\ref{sec:timing} we present results of fitting pulse
timing models to the data.  In \S\ref{sec:gravtheory} we use our
measurements to place constraints on theories of relativistic gravity.
In \S\ref{sec:massdisc} we discuss the pulsar mass and the evolution
of the binary system.  In \S\ref{sec:summary}, we summarize the major
points of the paper.

\section{Observations}\label{sec:obs}

\begin{figure}[b]
\plotone{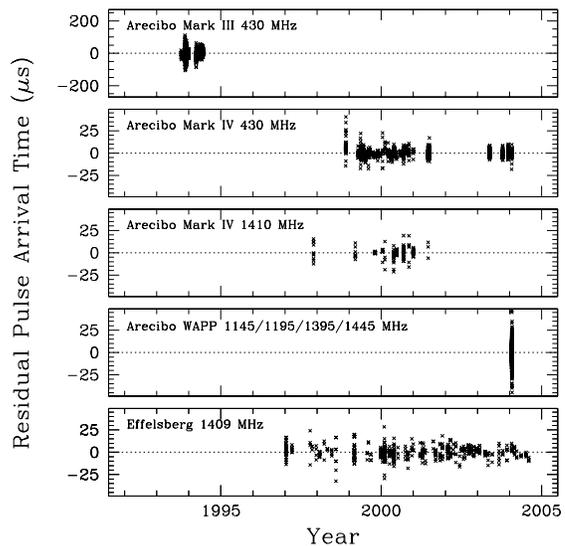}
\caption{Residual pulse arrival times after removing the
best fitting pulse timing model.  Points are segregated by
observatory, frequency, and data acquisition system.
Note the difference in vertical scales.
\label{fig:resid}}
\end{figure}

We observed \psr\ for more than ten years using the 305\,m telescope
at Arecibo and the 100\,m telescope at Effelsberg.  The bulk of the
Arecibo data were collected in campaigns of duration $\sim$1 week
over which all orbital phases were observed.  Such campaigns
were undertaken in November 1993, May 1999, May 2000, June/July 2001,
May
2003, and January 2004.  Additional data were collected at some other
epochs in 1993$-$1994 and 1999$-$2000 (Figure~\ref{fig:resid}).  
Effelsberg data were
less concentrated, with observations spread evenly over several years
and a single major campaign in February 2000.  The two types of data
complement each other, with campaign data especially useful for
measuring orbital elements, and with less concentrated observations
important for determining pulse spin-down and astrometric parameters.
Ultimately, all 5035 pulse arrival times from both observatories were
combined and simultaneously fit to pulsar timing 
models~(\S\ref{sec:timing}).

\begin{deluxetable*}{ll}[t]
\tablewidth{0pt}
\tablecaption{Timing Model Parameters\tablenotemark{a}\label{tab:param}}
\tablehead{\multicolumn{2}{c}{Basic Timing Model (Three~Post-Keplerian Parameters)}}
\startdata
Ecliptic longitude, $\lambda$ \dotfill                                  & $116\fdg 33362028(2)$  \\
Ecliptic latitude, $\beta$ \dotfill                                     & $-2\fdg 807548(2)$ \\  
Proper motion in $\lambda$, $\mu_\lambda=\cos\beta(d\lambda/dt)$ (mas/yr) \dotfill & $-$0.35(3) \\
Proper motion in $\beta$, $\mu_\beta$ (mas/yr) \dotfill                 & $-$6(2) \\  
Parallax (mas), $\pi$ \dotfill                                          & 1.6(8) \\ 
Rotation frequency, $\nu_0$ (s$^{-1}$) \dotfill                         & 287.457858630106(2)\\
Rotation frequency derivative, $\nu_1$ (s$^{-2}$)\tablenotemark{b} \dotfill  & $-6.4337(4)\times 10^{-16}$ \\
Epoch, $t_0$ (MJD [TDB]) \dotfill                                       & 51800.0 \\
Dispersion measure, DM$_0$ (pc\,cm$^{-3}$)\tablenotemark{c} \dotfill                     & 30.2489(3) \\
Dispersion measure derivative, DM$_1$ (pc\,cm$^{-3}$\,yr$^{-1}$) \dots  & $-0.00017(1)$ \\
Orbital period, $P_b$ (days)\tablenotemark{b} \dotfill                  & 0.263144266723(5) \\
Projected semi-major axis, $x$ (lt-s) \dotfill                          & 0.3966127(6) \\
Eccentricity, $e$ \dotfill                                              & $5(11)\times 10^{-7}$\\
Time of ascending node, $t_{\rm asc}$ (MJD [TDB]) \dotfill              & 51800.21573411(2) \\
Orbital period derivative, $\dot{P_b}$ (unitless) \dotfill              & $-6.2(8)\times 10^{-14}$ \\
Shapiro parameters, $r$ and $s$ \dotfill                                & see Figure \ref{fig:shapirors} \\

\cutinhead{General Relativistic Timing Model (Two Post-Keplerian Parameters)}

Cosine of inclination angle, $\cos i$\dotfill                       & $0.41^{+0.11}_{-0.07}$ \\
Pulsar mass, $m_1$ (M$_\odot$)\dotfill                              & 2.1(2)    \\
Secondary mass, $m_2$ (M$_\odot$)\dotfill                           & 0.191(15) \\

\enddata
\tablenotetext{a}{Figures in parentheses are 68\% confidence uncertainties in the last digit quoted.}
\tablenotetext{b}{Observed value, not corrected for acceleration biases;
see Table~\ref{tab:accel}.}
\tablenotetext{c}{Formal uncertainty in the timing fit.  No attempts were made to correct for pulse shape evolution over frequency.}
\end{deluxetable*}

\subsection{Arecibo Observations}

Observations at Arecibo in 1993 and 1994 are described in \cite{lzc95}.
The Princeton Mark III data acquisition system collected signals
across a 8\,MHz passband at 430\,MHz using a 32-channel filter bank
spectrometer with 100$\,\mu$ time constants.
Observations from 1997 to 2004 employed the Princeton Mark IV
coherent dedispersion system \citep{sst+00}.  Most observations were 
made at 430\,MHz with
a 5\,MHz passband, but some were made at 1410\,MHz using a 10\,MHz
passband.  Observations typically lasted 29 minutes and were analyzed
in blocks of 190\,s.

The January 2004 Arecibo campaign included extensive observations at radio
frequencies 1120 to 1470\,MHz.  These data were collected with four
Wideband Arecibo Pulsar Processors (WAPPs).  Each WAPP calculated
autocorrelations of 192 lags across a 50 MHz passband in each sense of
polarization.  Autocorrelations were accumulated for 32~$\mu$s and
then summed into folded pulse profiles with 256 bins across the pulse
period.  The four WAPP passbands were centered at 1145, 1195, 1395, and
1445\,MHz.  Data from each of the four passbands were reduced
independently.

Each data acquisition system folded the signals modulo the pulse
period using a precomputed ephemeris.  
A pulse time of arrival (TOA) was calculated from each
data block by fitting a high quality template profile to the data profile
and adding the phase offsets thus measured to the observation start
times.  A suitable number of pulse periods was added to the TOA
to make it fall near the center of the observation.
Start times were referenced to the
observatory time standard and later corrected to UTC and, ultimately,
TT(BIPM03).

\subsection{Effelsberg Observations}

At Effelsberg, timing data of PSR J0751+1807 have been collected since
1997, with observations made approximately once a month using a
1300$-$1700~MHz tunable HEMT receiver at a center frequency of
1409~MHz.  All observations
employed the Effelsberg--Berkeley Pulsar Processor (EBPP) as the data
acquisition system, correcting the dispersion smearing of the signal
using coherent dedispersion \citep{bdz+97}.

In total power mode, the EBPP provided 32 channels across 
a total of 56~MHz for each sense of circular polarization.
The output signals of each channel
were fed into digital dedisperser boards for coherent on-line
dedispersion and were synchronously folded at the pulse period over 
7~min integration times.  The signal-to-noise ratio varied between 3 and 10 
for the individual integrations, depending on interstellar scintillation.
As with the Arecibo data, TOAs were calculated by fitting
the data to high quality template profiles and calculating
the phase offset relative to the observation start times.
Start times were referenced to the
observatory hydrogen maser clock, corrected to UTC(NIST) using the
signals from the Global Positioning System (GPS), and finally
corrected to TT(BIPM03).

\section{Timing analysis}\label{sec:timing}

\subsection{Basic Timing Model (Three Post-Keplerian Parameters)}
\label{sec:basictiming}

The TOA measurements were fit to pulse timing models using the {\sc
tempo} software package\footnote{http://pulsar.princeton.edu/tempo}.
The basic timing model had 23 parameters and 5012 degrees of freedom.
The results of the fit is summarized in Table~\ref{tab:param}, and
residual arrival times after removing the timing model are plotted in
Figure~\ref{fig:resid}.  Pulsar rotation was parameterized by spin
frequency, $\nu_0$, and its derivative, $\nu_1$.  Earth motion was
modeled by the JPL DE405 Ephemeris \citep{sta98b}.  To minimize
covariances between components of position and proper motion,
astrometric results are presented in ecliptic coordinates, calculated
by rotating the coordinates of the solar system ephemeris about the
$x$ axis by the obliquity of the Earth, $\epsilon_0=84381\farcs 412$.
Arbitrary time offsets were fit between data sets collected at
different observatories, at different frequencies, or with different
equipment.  Interstellar dispersion was modeled with a dispersion
measure, DM, changing linearly with time, ${\rm DM}(t)={\rm DM}_0 +
{\rm DM}_1t$, while dispersion within the solar system was calculated
using an analytical model of the solar wind (\S\ref{sec:solarwind}).

Orbital motion was calculated using the {\sc ``ell1''} relativistic
orbital model \citep[Appendix A of][]{lcw+01}.  A combination of five
Keplerian and three post-Keplerian orbital elements were fit to the
data.  The Keplerian orbital elements were the orbital period, $P_b$;
the semi-major axis projected into the line of sight, $x=(a_1 \sin i)/c$,
where $i$ is the inclination and $c$ is the speed of light; the time of
passage through the ascending node, $t_{\rm asc}$; and two
Laplace-Lagrange parameters, $\eta$ and $\kappa$, which parameterize
the eccentricity.  Since neither of these were statistically
significant, we report total eccentricity,
$e=(\eta^2+\kappa^2)^{1/2}$; the value in Table~\ref{tab:param} should
be taken as an upper limit. 

The three post-Keplerian parameters were the time rate of change
of orbital period, $\dot{P_b}$, and the shape and range parameters
of Shapiro delay, $s$ and $r$.  The latter are defined
by the perturbation in arrival time, $\Delta t$, imposed by Shapiro
delay;  for a nearly circular orbit, this is
\begin{equation}
\Delta t = -2r \ln\{1-s \sin [(2\pi/P_b)(t-t_{\rm asc})]\},
\label{eqn:shapiro}
\end{equation}
where $t$ is the pulse arrival time.

\begin{figure}[b]
\plotone{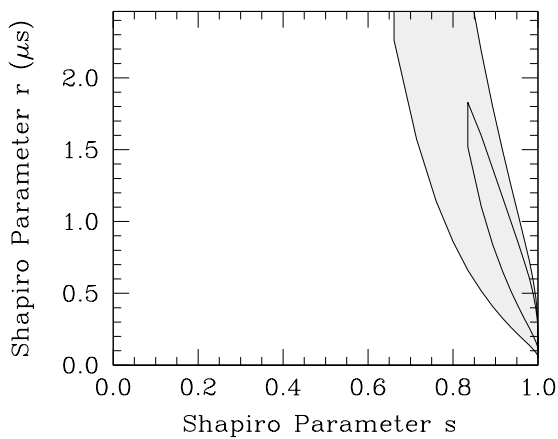}
\caption{Constraints on Shapiro delay parameters $r$ and $s$
from the basic timing analysis.  Inner and outer contours
delimit 68\% and 95\% confidence regions.}
\label{fig:shapirors}
\end{figure}

The best fit parameters are given in Table~\ref{tab:param} under the
heading ``Basic Timing Model.'' The most important new measurement is
the relativistic period decay rate,
\begin{equation}
\dot{P_b}=-(6.2\pm0.8)\times 10^{-14},
\label{eqn:pbdotmeas}
\end{equation}
discussed further below.  The two Shapiro parameters, $r$ and $s$, are
highly covariant, and the confidence regions are non-elliptical.  The
values allowed by the timing data are shown in Figure~\ref{fig:shapirors}.

\subsection{The Basic Timing Model under General Relativity}
\label{sec:basictiming2}

The basic timing model provides an excellent fit to the data without
reference to a specific theory of gravitation.  The implications of
this for gravitation theory are discussed below
(\S\ref{sec:gravtheory}).  Here, these results are interpreted in
terms of general relativity.

Under general relativity, the post-Keplerian parameters
are related to the masses of the pulsar, $m_1$, and the secondary star, $m_2$,
and the inclination of the orbit:
\begin{eqnarray}
  \left(\dot P_b\right)_{\rm GR}  & = & -\left(\!\frac{192\pi}{5}\!\right)\!
  \left(\!\frac{2\pi}{P_b}\!\right)^{\!5/3}\!
  \left(\!1\!+\!\frac{73}{24} e^2\!+\!\frac{37}{96} e^4\!\right)
\nonumber\\
 & & \hspace*{16pt} \times
  \frac{1}{(1\!-\!e^2)^{7/2}}\,T_\odot^{5/3}\,\frac{m_1\, m_2\,}{(m_1\!+\!m_2)^{1/3}},
\label{eqn:pbdot}
\end{eqnarray}
\vspace*{2pt}
\begin{equation}
r = T_\odot\, m_2,
\label{eqn:shapiror}
\end{equation}
and
\begin{equation}
s = \sin i,
\label{eqn:shapiros}
\end{equation}
where masses are in solar units and
$T_{\odot}\!=\!GM_\odot/c^3\!=\!4.925490947\!\times\!10^{-6}\,$s.
The masses and inclination are related by the Keplerian mass function,
\begin{equation}
f_1\equiv\frac{(m_2 \sin i)^3}{(m_1+m_2)^2}
        = \frac{x^3}{T_\odot}\left(\frac{2\pi}{P_b}\right)^2.
\label{eqn:f1}
\end{equation}

\begin{figure*}[t]
\plottwo{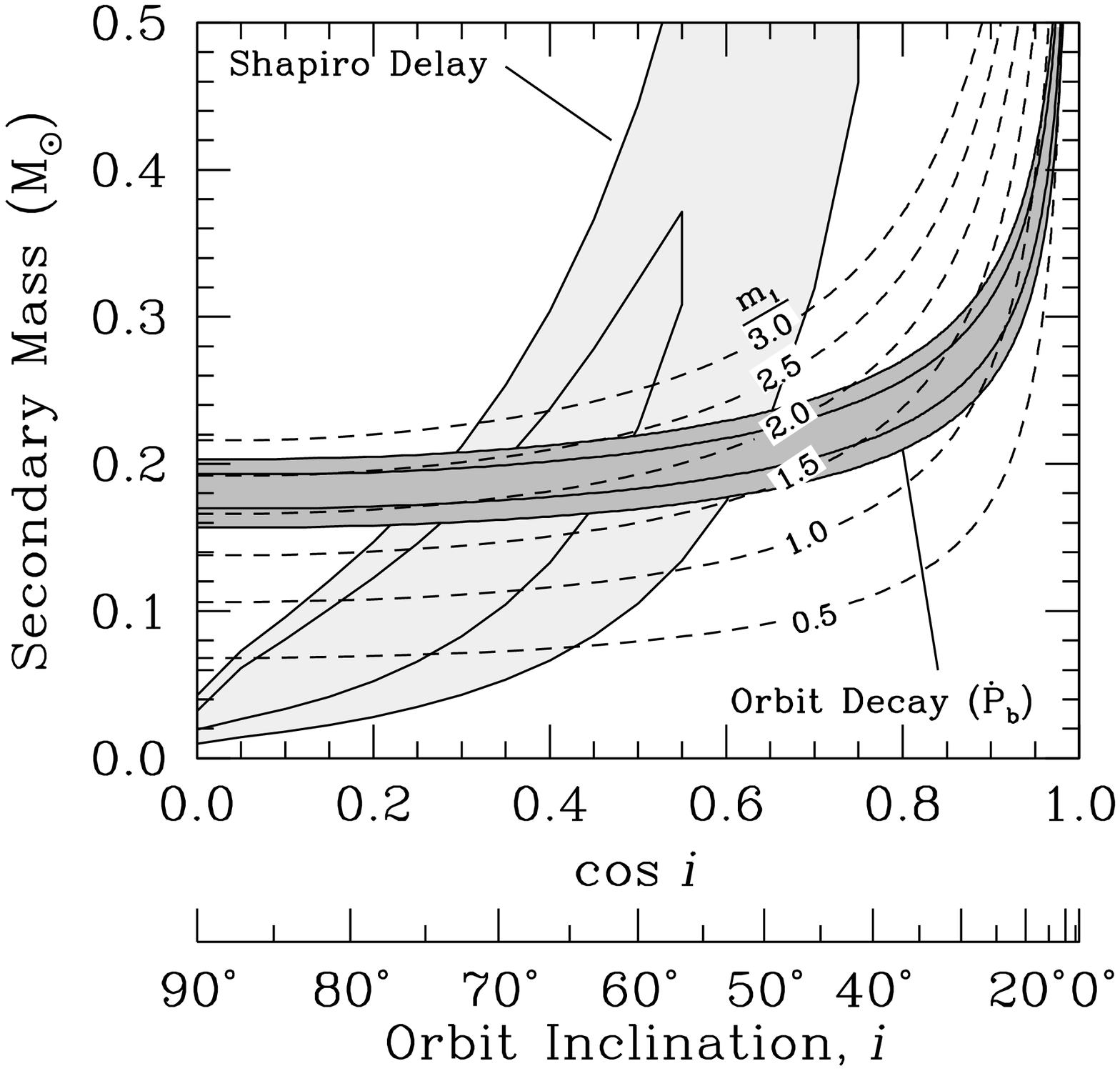}{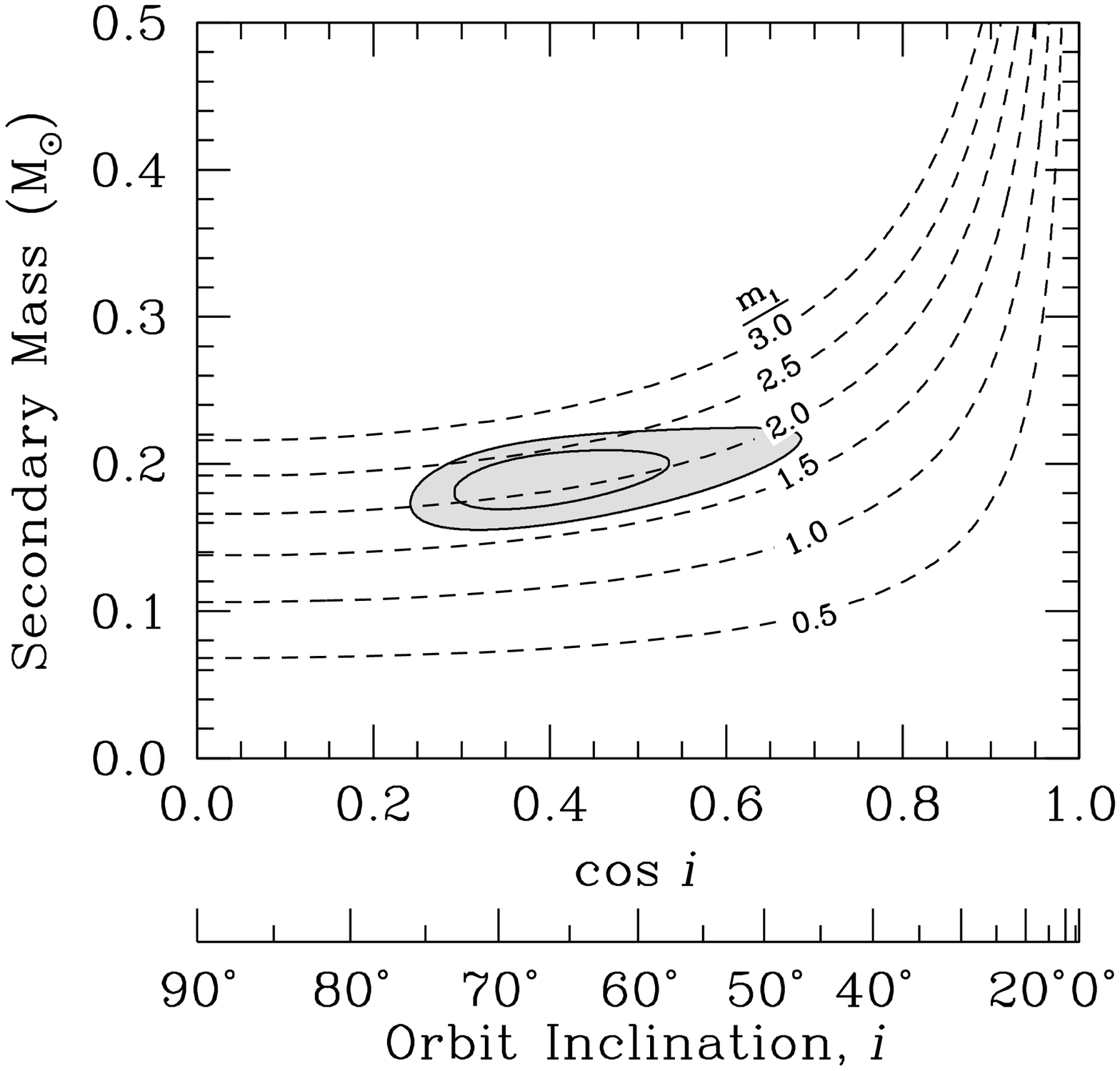}
\caption{Constraints on $\cos i$ and $m_2$.
Dashed lines indicate values of $m_1$ according to equation~\ref{eqn:f1}.  
(a) {\it Left plot:}  Constraints from the
basic timing model, with three post-Keplerian
parameters (orbital decay and two Shapiro delay parameters),
cast into inclination and mass values via equations~\ref{eqn:pbdot}
through ~\ref{eqn:f1}.
(b) {\it Right plot:} Constraints from the
general relativistic timing model, with two
post-Keplerian parameters (inclination and secondary mass).
In each plot, inner and outer contours correspond to 68\% and 95\% confidence
limits.  
\label{fig:cosim2}}
\end{figure*}

Under general relativity, the measured values of orbital decay
and Shapiro delay 
yield the constraints on $m_1$, $m_2$, and $i$ shown in
Figure~\ref{fig:cosim2}a.  
Evidently, meaningful constraints on the individual masses can
be attained only by combining the Shapiro delay and orbit
decay measurements.  Given this circumstance, it is useful to
directly incorporate general relativity into the timing analysis
and to explicitly consider $m_1$ and $m_2$ as independent variables.

\subsection{General Relativistic Timing Model (Two~Post-Keplerian
Parameters);
Stellar Masses}
\label{sec:grtiming}

\begin{figure}[b]
\plotone{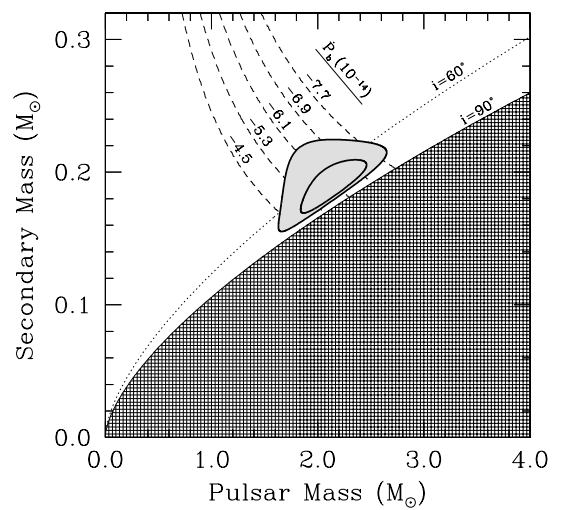}
\caption{Constraints on pulsar and secondary masses from the general
relativistic timing model.
Confidence limits of 68\% and 95\% are shown.  These are the
same constraints as the right plot of
figure~\ref{fig:cosim2}, cast into a different parameterization.
The shaded region in the lower left is disallowed
by the Keplerian mass function.  Dashed lines show constraints from
$\dot{P_b}$ alone.   A dotted line indicates inclination $i=60^{\circ}$.
\label{fig:m1m2}}
\end{figure}

We undertook a series of timing analyses in which general relativity
was assumed to hold, so that there were only two independent
post-Keplerian parameters instead of the three parameters of the basic
timing model.  This timing model had a total of 22 parameters and
5013 degrees of freedom.
We chose the independent variables to be $\cos i$ and
$m_2$. We analyzed a uniform grid of values of these quantities,
restricting $\cos i$ to be between 0 and 1, and $m_2$ to be between 0
and 0.5\,\msun.  For each value of $\cos i$ and $m_2$, the values of
$\dot{P_b}$, $m_1$, $r$, and $s$ were calculated using equations
\ref{eqn:pbdot} through \ref{eqn:f1}. 
 The TOAs were fit to a timing
solution with these parameters held fixed but all other parameters
allowed to vary.  For each combination of $\cos i$ and $m_2$, the
statistic $\Delta\chi^2=\chi^2-\chi^2_{\rm min}$, of the best fit
timing solution was recorded, where $\chi^2$ is the
goodness-of-fit of a particular timing solution and $\chi^2_{\rm min}$
is the global minimum from all grid points.  Fits with acceptable
values of $\Delta\chi^2$ at the 68\% and 95\% confidence levels are
shown in Figures~\ref{fig:cosim2}b and \ref{fig:m1m2}.

The values of 
$\cos i$, $m_1$, and $m_2$ and their uncertainties,
were calculated followed the procedure outlined in Appendix A of
\citet{sna+02}.
Essentially this was a Bayesian analysis with uniform
priors in $\cos i$ and $m_2$.
A probability was assigned to each grid point
based on its $\Delta\chi^2$.  After suitable
normalization, the probabilities of all
points associated with a given range of $\cos i$
(or $m_1$ or $m_2$) were summed to calculate a probability distribution
function.  The confidence intervals derived from these
probability distributions were:
\begin{equation}
\cos i = \left\{
\begin{array}{ll}
0.41^{+0.11}_{-0.07} & \mbox{(68\% confidence)} \\[4pt]
0.41^{+0.27}_{-0.13} & \mbox{(95\% confidence),}  \\
\end{array}
\right.
\end{equation}
\begin{equation}
m_1 = \left\{
\begin{array}{r@{}l@{\,}ll}
2.1 & \pm 0.2        & \msun       & \mbox{(68\% confidence)} \\[3pt]
2.1 & ^{+0.4}_{-0.5} & \msun       & \mbox{(95\% confidence),} \\
\end{array}
\right.
\end{equation}
and 
\begin{equation}
m_2 = \left\{
\begin{array}{r@{}l@{\,}ll}
0.191 & \pm 0.015           & \msun & \mbox{(68\% confidence)} \\[3pt]
0.191 & ^{+0.033}_{-0.029} & \msun & \mbox{(95\% confidence).} \\
\end{array}
\right.
\end{equation}

\subsection{Lower Limit on Pulsar Mass from $\dot{P_b}$ Alone}
\label{sec:pbdotmass}

While there is no reason to doubt the detection of Shapiro delay, it
is worth noting that the orbit decay measurement alone
provides some evidence for a very massive neutron
star.  We calculated a probability distribution function for $m_1$ 
assuming a uniform {\it a priori} distribution for $\cos i$, appropriate
for randomly oriented orbits, and assuming $\dot{P_b}$ is drawn from
the Gaussian distribution implied by its measurement uncertainty.  The
distribution of $m_1$ calculated under these assumptions gives lower
limits $m_1\!>\!1.75\,\msun$ (68\% confidence) and
$m_1\!>\!0.88\,\msun$ (95\% confidence).  

\subsection{Solar Wind}\label{sec:solarwind}

Radio signals are dispersed by electrons in the interstellar medium
and in the solar wind.  Because the solar wind is variable and
unpredictable, its contribution can degrade the accuracy of pulsar
timing models.  The problem is particularly acute for \psr\ because of
its low ecliptic latitude.  Dispersion  by the solar wind imposed
delays of up to $\sim$12\,$\mu$s in the TOAs in our data set.  
We excluded all observations for which the line of sight to the
pulsar passed within $15^\circ$ of the sun.
We found that changing this cutoff 
to $30^\circ$ or to $0^\circ$ had little impact on our results.

In
principle, observations at multiple radio frequencies at every epoch
would allow the dispersion to be measured and removed; in practice, it
is difficult to measure high precision TOAs at two or more widely
spaced frequencies, so it becomes necessary to use an analytic model
of the solar wind and average the results over many epochs.
We modeled the electron density in the solar wind as $n_e(r)=n_0(1
\mbox{\sc AU}/r)^2$, where $r$ is the distance to the sun and $n_0$ is
the electron density at 1\,{\sc AU}.  We found the best fits to the
data had $n_0=9.6\pm 3.0$\,pc\,cm$^{-3}$.  Within this uncertainty
range, the particular value of $n_0$
used had
little impact in the rotation and binary parameters derived in the
timing fit;  we used a fixed value of $n_0=9.6$\,cm$^{-3}$ to
calculate the rotation and binary parameters in Table~\ref{tab:param}.

The pulsar's astrometric parameters---position, proper motion, and 
parallax---exhibit high covariances with $n_0$ and with time changes in $n_0$.  
To estimate values and uncertainties for these parameters, we fit a grid
of timing solutions with $n_0$ between 6.6 and 12.6\,pc\,cm$^{-3}$,
and with the time derivative of $n_0$ between $-$1.5 and
+1.5\,pc\,cm$^{-3}$\,yr$^{-1}$.  We use the extreme values from
these timing fits to calculate the uncertainties on 
position, proper motion, and parallax  given in Table~\ref{tab:param}.

Observations of another pulsar, J1713+0747, over a similar period of
time found a marginally smaller solar wind electron density,
$n_0=5\pm4$\,pc\,cm$^{-3}$ \citep{sns+05}.  Despite differing by a
factor of two, the $n_0$ values from the two pulsars are in
statistical agreement.

\section{Probing Strong Field Gravity}\label{sec:gravtheory}
\nopagebreak

Pulsars are well established testbeds for relativity.  Observations of
the neutron star--neutron star binary
PSR~B1913+16 have established that its orbit decays at the rate
predicted by general relativity within 0.3\% \citep{tw89,wt03}.
However, this test of relativity is of limited use for constraining
violations of the strong equivalence principle (SEP) because of the
similar self-energies of the two neutron stars.  More useful for
probing this aspect of gravitation are neutron star--white dwarf
binaries, in which the radically different self-energies of the two
stars would generate excess gravitational wave energy loss under
SEP-violating theories.  A succinct review is given by \cite{arz03}
\citep[see also][]{sta03,wz89,gol92,lcw+01,gw02}.

The observed change in the orbital period of \psr\ is
$(\dot{P_b}/P_b)_{\rm obs}=(-2.7\pm0.4)\times 10^{-18}\,{\rm s}^{-1}$.
We consider five possible mechanisms for generating changes in the
orbital period: acceleration of the binary system relative to the
Earth; changes in the gravitational constant, i.e., nonzero $\dot{G}$;
changes in the masses of the component stars; energy
loss from the binary due to quadrupole gravitational radiation, as in
general relativity; and energy loss due to dipole radiation, as in
SEP-violating theories. Contributions from tidal interactions between 
the two stars are likely not to be important \citep[e.g.,][]{sb76}.
 The observed decay rate is the sum of the
five contributions:
\begin{eqnarray}
\left(\frac{\dot{P_b}}{P_b}\right)_{\!\rm obs} & =  &
 \left(\frac{\dot{P_b}}{P_b}\right)_{\!A} 
+\left(\frac{\dot{P_b}}{P_b}\right)_{\!\dot{G}} 
+\left(\frac{\dot{P_b}}{P_b}\right)_{\!\dot{m}}
\nonumber\\
& &
\hspace*{36pt}
+\left(\frac{\dot{P_b}}{P_b}\right)_{\!Q} 
+\left(\frac{\dot{P_b}}{P_b}\right)_{\!D}.
\label{eqn:pbdotterms}
\end{eqnarray}
We address each of these terms in turn.

The first term, acceleration biases,
arises from the proper motion of the pulsar, acceleration toward
the Galactic plane ($z$-acceleration), and Galactic rotation
\cite[see][for details]{dt91}.  The biases for \psr\ are listed in
Table~\ref{tab:accel}.  These values were calculated using the
measured proper motion of 6\,mas\,yr$^{-1}$, and using a distance of
1.15\,kpc, derived from the dispersion measure of the pulsar using the
NE2001 galactic electron density model \citep{cl02}. 
To be conservative, we assign an uncertainty equal to the total
bias (i.e., 100\% uncertainty).  The bias of $\dot{P_b}$
is much smaller than its measurement
uncertainty and essentially negligible.\footnote{The same phenomena
shift the pulse period derivative, $\dot{P}=\nu_1/\nu_0^2$, a few percent
away from its intrinsic value, but this is of
no practical consequence.}

The second term in Equation~\ref{eqn:pbdotterms}, due to $\dot{G}$, can be 
shown negligible by
appeal to other binary pulsars.  The change in orbital period due to
nonzero $\dot{G}$ is approximately $(\dot{P_b}/P_b)_{\dot{G}} \simeq
-2(\dot{G}/G).$ This expression neglects the effects of $\dot{G}$ on
the energy content of the stars themselves;  a more precise expression
is given in \cite{nor90}, but is not necessary for the purposes of the
present paper.  Because the same expression for
$(\dot{P_b}/P_b)_{\dot{G}}$ holds for {\it any} binary pulsar system,
the binary with the lowest value of this expression sets an upper
limit on it for {\it all} pulsars.  The best limit comes from
PSR~J1713+0737, for which $\dot{P_b}/P_b\leq 1\times 10^{-19}\,{\rm
s}^{-1}$  \citep{sns+05}, more than an order of magnitude smaller than
our value for \psr.  A comparable limit on $(\dot{P_b}/P_b)_{\dot{G}}$
can be set using limits on $\dot{G}$ from lunar laser ranging
measurements, which have found $\dot{G}/G=(1.3\pm 2.9)\times
10^{-20}\,{\rm s}^{-1}$ \citep{wtb04}.  In any case, $\dot{G}$ effects
are unimportant for \psr.

\begin{deluxetable}{@{\hspace*{18pt}}lrr}[t]
\tablecaption{Biases of pulsar and orbital period derivatives\label{tab:accel}}
\tablehead{Quantity  & \multicolumn{1}{c}{$\dot{P}$} & \multicolumn{1}{c}{$\dot{P_b}$} }
\startdata
\multicolumn{3}{l}{\it Measurement \dots} \\
                       &      $7.7860\times 10^{-21}$ &     $-6.2\times 10^{-14}$ \\
Uncertainty            &  $\pm 0.0005\times 10^{-21}$ &  $\pm 0.8\times 10^{-14}$ \\[3pt]
\multicolumn{3}{l}{\it Acceleration biases \dots} \\
Proper motion          &  $0.35\phn\phn\times 10^{-21}$ &  $0.2\times 10^{-14}$ \\
$z$-acceleration       & $-0.19\phn\phn\times 10^{-21}$ & $-0.1\times 10^{-14}$  \\
Galactic rotation      &  $0.19\phn\phn\times 10^{-21}$ &  $0.1\times 10^{-14}$ \\[3pt]
\multicolumn{3}{l}{\it Intrinsic value \dots} \\
Measurement$-$Bias     &  $7.44\phn\phn\times 10^{-21}$ & $-6.4\times 10^{-14}$ \\
Uncertainty            &  $\pm 0.4\phn\phn\phn\times 10^{-21}$ & $\pm0.9\times 10^{-14}$ \\
\enddata
\end{deluxetable}

The third term arises if one of the component stars is losing mass
\citep{eh75,lbd+05}.  First, we consider mass loss from
the pulsar.  The pulsar's measured spin down rate implies
that it is losing energy at a rate $\dot{E}=4\pi^2\nu_0\nu_1 I_1,$
where $I_1$ is the moment of inertia of the pulsar.  Assuming the
pulsar outflow is relativistic, this energy loss
rate is equivalent to a mass loss rate of
$\dot{m_1}=\dot{E}/c^2=4\pi^2\nu_0\nu_1 I_1/c^2=-4\times
10^{-21}\,\msun\,{\rm s}^{-1}(I_1/10^{45}\,{\rm g\,cm})$.  The
resulting change in the orbital period is
$(\dot{P_b}/P_b)_{\!\dot{m}}=-2\dot{m_1}
(m_1+m_2)^{-1}\simeq -4\times 10^{-21}\,{\rm s}^{-1}$,
depending on the precise values of $m_1$, $m_2$, and $I_1$.  This is
three orders of magnitude smaller than the measured value and, so
mass loss can be neglected.  

Next, we consider mass loss from the white dwarf companion.
I.~Wasserman (private communication) has pointed out that mass flow
from the white dwarf could arise from irradiation of the white dwarf
by the pulsar.  If such an outflow were directed predominantly opposite
the motion of the white dwarf, it would induce a negative $\dot{P_b}$.
Given the geometry of the \psr\ system, a white dwarf of low mass might
capture sufficient the pulsar flux to produce such an outflow,
although there is no reason to expect the flow to be collimated tangential
to the orbit.  The presence of an ionized outflow would give rise to
variations in the dispersion measure of the pulsar over the course of
its orbit and, depending on the orbital geometry, might even cause
eclipses.  However, \psr\ exhibits neither eclipses nor orbital
variability in its dispersion measure.  To search for the latter, we
analyzed multi-frequency observations made in January 2004 by dividing the
orbit into ten equal length sections and separately measuring the
dispersion measure in each section.  We found no significant variation
in dispersion measure over the course of the orbit, with a
conservative upper bound of 
$\Delta{\rm DM}<4\times 10^{-4}\,{\rm pc}\,{\rm cm}^{-3}
=1\times10^{15}\,{\rm cm}^{-2}$.  This is strong, but not
definitive, evidence against an outflow; it is possible to imagine
scenarios with small orbital inclinations in which the pulsar
irradiation directs the outflow away from our line of sight.  Optical
observations of the companion find it to be relatively cool, further
evidence that a significant wind is unlikely (van Kerkwijk et al.
2005;  M. van Kerkwijk, private communication)\nocite{vbjj05}.

The expressions for $(\dot{P_b}/P_b)_Q$ and $(\dot{P_b}/P_b)_D$ in
Equation~\ref{eqn:pbdotterms} depend
on gravitational theory.  In general relativity, $(\dot{P_b}/P_b)_Q$
is given by Equation~\ref{eqn:pbdot}, while $(\dot{P_b}/P_b)_D=0$.  In
some SEP-violating theories, the quadrupole term equals the general
relativistic expression, while the dipole contribution 
has the form
\begin{equation}
\left(\frac{\dot{P_b}}{P_b}\right)_{\!D} = 
-\left(\!\frac{2\pi}{P_b}\!\right)^{\! 2}
T_{\odot}
\left(\!\frac{G_*}{G}\!\right)
\frac{m_1m_2}{m_1+m_2}
(\alpha_{c1}-\alpha_{c2})^2.
\label{eqn:dipole}
\end{equation}
where $\alpha_{c1}$ and $\alpha_{c2}$ are the couplings
of the pulsar and the secondary star, respectively, to the scalar field,
and $G_*$ is the ``bare'' gravitational constant.

To measure or constrain the dipole term $(\dot{P_b}/P_b)_D$ using
Equation~\ref{eqn:pbdotterms}, it is necessary to know the quadrupole
term $(\dot{P_b}/P_b)_Q$, which in turn requires knowing $m_1$ and
$m_2$ by means other than the measured $\dot{P_b}$.  For \psr, there
are no independent high precision measurements of $m_1$ and $m_2$, so we
must consider all combinations of $m_1$ and $m_2$ which fall within the
Shapiro delay 95\% confidence contour (Fig~\ref{fig:cosim2}a) and have
reasonable neutron star masses, $1\,\msun\!<\!m_1\!<\!3\,\msun$ \citep{lp04}.
For these masses, the predicted general relativistic values of
$(\dot{P_b}/P_b)_Q$ range from $-1.1\times 10^{-18}\,{\rm s}^{-1}$ to $-1.4\times
10^{-17}\,{\rm s}^{-1}$.  The largest negative allowed value of 
$(\dot{P_b}/P_b)_D=(\dot{P_b}/P_b)_{\rm obs}-(\dot{P_b}/P_b)_Q$,
is $-4.1\times 10^{-18}\,{\rm s}^{-1}$ (95\% confidence).  This arises at the
smallest values of $m_1$ and $m_2$ that fall within the constraints, and
corresponds to a difference in coupling strengths
\begin{equation}
(\alpha_{c1}-\alpha_{c2})^2 < 7\times 10^{-5}.
\label{eqn:alpha}
\end{equation}
This improves by a factor of a few on previously published upper
limits from PSRs~B0655+64 and J1012+5307 \citep{arz03,lcw+01}.  A
result comparable to it can be derived from the
measurements of the PSR~J1141$-$6545 orbit presented in \cite{bokh03}.
In fact, in generalized tensor-scalar theories of gravity, 
the latter pulsar is likely to be the most constraining among known
pulsar--white dwarf binaries.   Its eccentric orbit gives rise to
several relativistic phenomena, allowing its stellar masses to be
measured independently of $\dot{P_b}$ \citep{esp04}.

In Brans-Dicke theory, the coupling to star $i$ is
$\alpha_{ci}=2s_i(2+\omega_{\rm BD})^{-1}$, where $s_i=-\partial\ln
m_i/\partial\ln G$, the ``sensitivity'' of star $i$, is the change in
its binding energy as a function of $G$, and where $\omega_{\rm BD}$
is the Brans-Dicke coupling constant.  Estimates of $s_1$ range from
0.1 to 0.3, depending on the neutron star equation of state
\citep{wz89}.  The sensitivity of the white dwarf is negligible.  The
limit on coupling strengths (Eqn.~\ref{eqn:alpha}) places a lower limit
$w_{\rm BD}>1300(s_1/0.2)^2$ on the Brans-Dicke coupling constant.
This limit is higher than previous constraints from pulsar work, but
lower than the best limit attained by other means, $\omega_{\rm
BD}>40000$, from radio ranging to the Cassini probe \citep{bit03}.

\begin{figure}[b]
\plotone{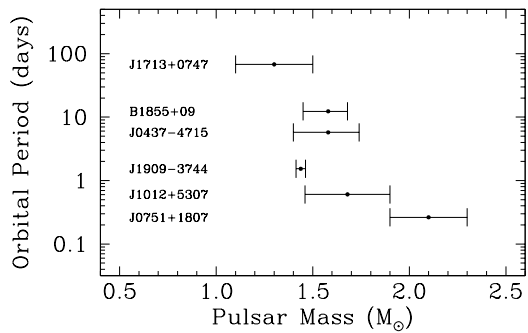}
\caption{Measured pulsar masses in circular pulsar--helium white dwarf binary
systems as a function of orbital period.  Data are from
this paper, \cite{jhb+05}, \cite{lcw+01}, \cite{vbb+01}, and \cite{nss05}, and 
references therein.
\label{fig:m1pb}}
\end{figure}

\section{Discussion}\label{sec:massdisc}

The mass of \psr\ is the largest measured for any pulsar.  As shown in
Figure~\ref{fig:m1pb}, pulsars in circular orbits with helium
white dwarf companions tend to have masses greater than the canonical
value of 1.35\,\msun.  This is in contrast to pulsars and secondary
stars in neutron star--neutron binaries, which fall within the range
1.18$-$1.44\,\msun\ (\S\ref{sec:intro}).  The relatively high masses of
pulsars in neutron star--white dwarf systems presumably result from
extended mass accretion during the late stages of their evolution.

An inverse correlation between orbital period and pulsar mass is
apparent in Figure~\ref{fig:m1pb}.  Any such relation is likely
to be complicated by the different evolutionary paths followed
by different systems.
The four systems with the longest orbital periods, 
those of PSRs J1713+0747, B1855+09, J0437$-$4715, and J1909$-$3744, 
are classical wide millisecond 
pulsar--helium white dwarf binaries, which underwent extended stable 
mass transfer.  In the library of evolutionary tracks calculated by 
\cite{prp02b}, such systems show an inverse correlation between orbital 
period and pulsar mass, although there is a wide spread in pulsar mass 
for a given orbital period.  In these calculations, it was assumed
that half of the mass lost by the secondary was accreted onto the
neutron star;  whether the mass accreted onto the neutron star is,
in fact, proportional to the mass lost by the secondary remains an
open question.

PSRs~J0751+1807 and J1012+5307, with their short
orbital periods, provide a challenge to binary evolution theories.
Evolution calculations tend not to produce systems with binary
periods close to the orbital period of \psr\ at the end of mass transfer.
Its orbital period
falls above the periods of ultracompact systems but below the
periods of standard pulsar--white dwarf binaries \citep[e.g.,][]{prp02b}.
\cite{esg01} are able to produce the properties of the \psr\ system
by invoking magnetic braking and heating of the secondary
by irradiation from the pulsar.  The latter effect drives mass loss
in the secondary, increasing the orbital separation and preventing the
binary from  shrinking into an ultracompact system.  
This picture is supported by optical observations, which find that the secondary lacks
a hydrogen envelope and that it has cooled rapidly
\citep{vbjj05}.

\section{Summary}\label{sec:summary}

The orbit of \psr\ decays at a rate of $\dot{P_b}=-(6.4\pm 0.9)\times
10^{-14}$.  This is in line with the expected value from general
relativity.  Combined with Shapiro delay measurements, it implies the
pulsar and secondary star masses are $2.1\pm 0.2$\,\msun\ and
$0.191\pm 0.15$\,\msun, respectively.  The mass of \psr\ is the
largest recorded for a pulsar, and it may imply greater mass is
transferred in tighter low mass neutron star binary systems than in
wider systems.

The maximum mass attainable by a neutron star depends on the
stars composition and the equation of state of nuclear matter.
\cite{lp01} calculated neutron star mass--radius relationships
for a number of plausible equations of state.  They found
that models with exotic components, such as pure quark
stars or mixed phases with kaon condensate or strange quark matter,
allow neutron stars to attain masses no higher than $\sim 2$\,\msun.
While the measurement of the \psr\ mass is nominally above this
limit, the measurement uncertainty is not yet small enough to draw firm
conclusions on this point.

The uncertainty in the measurement of $\dot{P_b}$ scales with
observation span to the $-$2.5 power for uniformly sampled data.  The
precision of this measurement can, therefore, be increased
dramatically with a relatively small number of observations over the
next few years.  The Shapiro delay measurement, because it has no such
time dependence, will improve more slowly, and hence uncertainty in
the Shapiro delay (and hence in the inclination) will ultimately
dominate the uncertainty in the mass measurement.

\acknowledgements

We thank I. Wasserman for useful discussions.
We thank D. Backer, A. Lommen, and K. Xilouris for collaborating
in the collection of portions of the Arecibo data.
The Arecibo Observatory is a facility of the National Astronomy
and Ionosphere Center, operated by Cornell University under a
cooperative agreement with the National Science Foundation.
DJN and JMC are supported by NSF grants AST-0206205 (to Princeton),
and AST-0206036 (to Cornell), respectively. 
IHS holds an NSERC UFA and is supported by a Discovery Grant.


\end{document}